\documentclass[conference]{IEEEtran}
\IEEEoverridecommandlockouts
\usepackage{cite}
\usepackage{amsmath,amssymb,amsfonts}
\usepackage{graphicx}
\usepackage{textcomp}
\usepackage{xcolor}
\usepackage{balance}  
\usepackage{subcaption}
\usepackage{amsmath}
\usepackage[utf8]{inputenc}
\usepackage{algorithm}
\usepackage[noend]{algpseudocode}
\usepackage{multirow}
\usepackage{array}

\newtheorem{problem}{Problem}
\newtheorem{theorem}{Theorem}[section]

\newtheorem{defn}[theorem]{Definition}

\def\BibTeX{{\rm B\kern-.05em{\sc i\kern-.025em b}\kern-.08em
    T\kern-.1667em\lower.7ex\hbox{E}\kern-.125emX}}
\begin{document}

\title{motif2vec: Motif Aware Node Representation Learning for Heterogeneous Networks}

\author{\IEEEauthorblockN{Manoj Reddy Dareddy*\thanks{*Work done as an intern at Visa Research.}}
\IEEEauthorblockA{\textit{University of California, Los Angeles}\\
Los Angeles, CA, USA\\
mdareddy@g.ucla.edu}
\and
\IEEEauthorblockN{Mahashweta Das}
\IEEEauthorblockA{\textit{Visa Research}\\
Palo Alto, CA, USA \\
mahdas@visa.com}
\and
\IEEEauthorblockN{Hao Yang}
\IEEEauthorblockA{\textit{Visa Research}\\
Palo Alto, CA, USA \\
haoyang@visa.com}
}

\maketitle

\begin{abstract}
Recent years have witnessed a surge of interest in machine learning on graphs and networks with applications ranging from vehicular network design to IoT traffic management to social network recommendations. Supervised machine learning tasks in networks such as node classification and link prediction require us to perform feature engineering that is known and agreed to be the key to success in applied machine learning. Research efforts dedicated to representation learning, especially representation learning using deep learning, has shown us ways to automatically learn relevant features from vast amounts of potentially noisy, raw data. However, most of the methods are not adequate to handle heterogeneous information networks which pretty much represents most real world data today. The methods cannot preserve the structure and semantic of multiple types of nodes and links well enough, capture higher-order heterogeneous connectivity patterns, and ensure coverage of nodes for which representations are generated. In this paper, we propose a novel efficient algorithm, {\em motif2vec} that learns node representations or embeddings for heterogeneous networks. Specifically, we leverage higher-order, recurring, and statistically significant network connectivity patterns in the form of motifs to transform the original graph to motif graph(s), conduct biased random walk to efficiently explore higher order neighborhoods, and then employ heterogeneous skip-gram model to generate the embeddings. Unlike previous efforts that uses different graph meta-structures to guide the random walk, we use graph motifs to transform the original network and preserve the heterogeneity. We evaluate the proposed algorithm on multiple real-world networks from diverse domains and against existing state-of-the-art methods on multi-class node classification and link prediction tasks, and demonstrate its consistent superiority over prior work. 
\end{abstract}


\textbf{{\em Author Terms $-$ } heterogeneous information networks, network embedding, network representation learning, feature learning, motifs}

\section{Introduction}
\label{sec:intro}

Recent years have witnessed a surge of interest in machine learning on graphs and networks with applications ranging from vehicular network design to IoT traffic management to drug discovery to social network recommendations. Graph-based data representation enables us to understand objects with respect to the neighboring world instead of just observing them in isolation. Thus, there is an increasing trend of representing data, that is not naturally connected, as graphs. Examples include item graph constructed from users’ behavior history that is originally sequential in nature~\cite{DBLP:conf/kdd/WangHZZZL18}, product review graph constructed from reviews written by users for stores~\cite{DBLP:conf/icdm/WangXLY11}, credit card fraud network constructed from fraudulent and non-fraudulent transaction activity data~\cite{DBLP:journals/dss/VlasselaerBCEAS15}, 
etc.          

Supervised machine learning tasks over nodes and links in networks\footnote{We use the term {\em network (nodes, links)} and {\em graph (vertices, edges)} interchangeably throughout the paper.} such as node classification and link prediction require us to perform feature engineering that is known and agreed to be the key to success in applied machine learning. However, feature engineering is challenging and tedious since the traditional process relies on domain knowledge, intuition, data manipulation, and manual intervention. Research efforts dedicated to representation learning, i.e., learning representations of the data that make it easier to extract useful information when training classifiers or other predictors, has shown us ways to automatically learn relevant features from vast amounts of potentially noisy, raw data. Of particular interest to the academic and industry research community has been representation learning using deep learning~\cite{DBLP:journals/pami/BengioCV13} that are formed by the composition of multiple non-linear transformations with the goal of yielding more useful representations. There has been a series of work over the past demi decade that focuses on graph node representation or graph embedding algorithms~\cite{DBLP:journals/kbs/GoyalF18}. The common goal of these works is to obtain a low-dimensional feature representation of each node of the graph such that the method is scalable and the vector representation preserves some structure and connectivity pattern between individual nodes in the graph. The graph embedding methods are broadly classified into three categories namely factorization based, random walk based, and deep learning based with applications in network compression, visualization, clustering, link prediction, and node classification~\cite{DBLP:journals/kbs/GoyalF18}. Among the three categories, random walk based graph embedding techniques have emerged to be the most popular since they help approximate many network properties, are useful when network is too large to measure in its entirety, and can work with partially observable network. The popular random-walk based methods include DeepWalk~\cite{DBLP:conf/kdd/PerozziAS14}, node2vec~\cite{DBLP:conf/kdd/GroverL16}, LINE~\cite{DBLP:conf/www/TangQWZYM15}, HARP~\cite{DBLP:conf/aaai/ChenPHS18}, etc.

However, most of these methods are designed for homogeneous networks and are inadequate to handle heterogeneous information networks, i.e., networks with multiple types of nodes and links, which pretty much represents most real world data today. Contemporary information networks like Facebook, DBLP, Yelp, Flickr, etc. contain multi-type interacting components. For example, social network Facebook has different types of objects (nodes) such as users, posts, photos as well as different kinds of associations (links) such as user-user friendship, person-photo tagging relationship, post-post replying relationship, etc. Researchers today acknowledge that heterogeneous networks fuse more information and support richer semantic representation of the real world~\cite{DBLP:journals/tkde/ShiLZSY17}\cite{DBLP:series/synthesis/2012Sun}. They also emphasize that data mining approaches designed for homogeneous graphs are not well-suited to handle heterogeneous graphs. For example, classification in homogeneous networks is traditionally done on objects of the same entity type, makes strong assumptions on the network structure, and assumes that data is independently and identically distributed (i.i.d.). Contrarily, classification in heterogeneous networks need to simultaneously classify multiple types of objects which may be organized arbitrarily and may violate the i.i.d assumption. Thus, there is an innate need to develop graph embedding methods for heterogeneous networks. 

Dong {\em et al.} formally introduced the problem and proposed a novel algorithmic solution metapath2vec~\cite{DBLP:conf/kdd/DongCS17} that leverages {\em metapath}, the most popular graph meta-structure for heterogeneous network mining~\cite{DBLP:series/synthesis/2012Sun}. A more recent work proposed metagraph2vec~\cite{DBLP:conf/pakdd/ZhangYZZ18} that leverages {\em metagraph} in order to capture richer structural contexts and semantics between distant nodes. Other heterogeneous network embedding methods include PTE~\cite{DBLP:conf/kdd/TangQM15} that is a semi-supervised representation learning method for text data; HNE~\cite{DBLP:conf/kdd/ChangHTQAH15} that learns representation for each modality of the network separately and then unifies them into a common space using linear transformations; LANE~\cite{DBLP:conf/wsdm/HuangLH17} that generates embeddings for attributed networks; and ASPEM~\cite{DBLP:conf/sdm/ShiGZK018} that captures the incompatibility in heterogeneous networks by decomposing the input graph into multiple aspects 
and learns embeddings independently for each aspect.
None of PTE, HNE, LANE, or ASPEM is aligned to the generic task of task-independent heterogeneous network embedding learning. The heterogeneity in PTE stems from links in a text network while the raw input belongs to the same object type; HNE works on a heterogeneous graph with image and text where the simultaneous interactions among multi-typed objects are decomposed into several scattered pairwise interactions in a single-typed network; LANE defines heterogeneity as diverse information sources (namely, network topology and node label information) that need to be jointly learnt; while ASPEM models heterogeneous network incompatibility and learns embeddings for each aspect independently. Among the related art, metapath2vec and metagraph2vec consider the general problem of learning node representations for heterogeneous networks. However, the methods cannot preserve the structure and semantics of multi-type nodes and links well enough, capture higher-order heterogeneous connectivity patterns, and ensure coverage of nodes for which representations are generated, as demonstrated in Section~\ref{sec:expt}.       
  
In this paper, we propose a novel efficient algorithm {\em motif2vec} that learns node representations or embeddings for heterogeneous information networks. Specifically, we leverage higher-order, recurring, and statistically significant network connectivity patterns in the form of {\em motifs} to learn higher quality embeddings. Motifs are one of the most common higher-order data structures for understanding complex networks and have been popularly recognized as fundamental units of network~\cite{DBLP:journals/science/BensonGL16}. It has been successfully used in many network mining tasks such as clustering~\cite{DBLP:conf/www/TsourakakisPM17}\cite{DBLP:conf/kdd/YinBLG17}, anomaly detection~\cite{DBLP:conf/allerton/Tsourakakis16}, and convolution~\cite{DBLP:journals/corr/abs-1711-05697}. However, no prior work has investigated the scope and impact of motifs in learning node embeddings for heterogeneous networks. Rossi {\em et al.} introduced the problem of higher-order network representation learning using motifs for homogeneous networks~\cite{DBLP:conf/www/RossiAK18}. But the method cannot be extended to handle heterogeneous networks. HONE~\cite{DBLP:conf/www/RossiAK18} does not combine the best of both worlds $-$ random walk based method that accounts for local neighborhood structure and motif-aware method that accounts for higher-order global network connectivity patterns, as we do. In addition, HONE (as well other existing methods) do not include the original network in the learning process, as we do. The latter ensures higher coverage of connected nodes.

Our algorithm motif2vec transforms the original graph to motif graph(s), conduct biased random walk to efficiently explore higher order neighborhoods, and then employ heterogeneous skip-gram model to generate the embeddings. Related efforts in heterogeneous network node embedding, namely, metapath2vec~\cite{DBLP:conf/kdd/DongCS17} and metagraph2vec~\cite{DBLP:conf/pakdd/ZhangYZZ18} are limited to only exploring neighborhoods, nodes, and links participating in the meta-structure of interest. motif2vec leverages motifs to transform the original graph to a motif representation and conduct regular random walks on the entire transformed graph. We evaluate our algorithm on multiple real-world networks from diverse domains and against existing state-of-the-art techniques on multiple machine learning tasks and demonstrate its consistent superiority over prior work. To summarize, we make the following contributions: 
\begin{itemize}
\vspace{-0.05in}
\item We propose {\em motif2vec}, an efficient and effective novel algorithm for representation learning in heterogeneous information networks. Specifically, we leverage higher-order, recurring, and statistically significant network connectivity patterns in the form of {\em motifs} to learn higher quality embeddings.
\item Our method preserves both local and global higher-order structural relationships as well as semantic correlations in a heterogeneous network. Unlike existing efforts, our method does not focus on refining the random walk to achieve the goal. Instead, we present a graph transformation method that enable us to capture sub-graph pattern significances.  
\item We empirically evaluate our algorithm for multiple heterogeneous network mining tasks, namely multi-class classification and link prediction on multiple real-world datasets from different domains and demonstrate its consistent superiority over state-of-the-art baselines.
\end{itemize}
\section{Preliminaries}

We introduce our problem definition and related concepts and notations before presenting our framework in Section~\ref{sec:framework}. 

\begin{defn}{\textbf{Heterogeneous Information Network:}}
A heterogeneous information network is defined as a directed graph G = (V, E, $T_V$, $T_E$) in which each node v $\in$ V is associated with mapping function $\phi$(v): V $\rightarrow$ $T_V$ and each link e $\in$ E is associated with mapping function $\varphi$(v): E $\rightarrow$ $T_E$. $T_V$ and $T_E$ denote the sets of node types and link types in G, $|T_V| > $ 1 and $|T_E| >$ 1.       
\end{defn}
    
\begin{figure}[t]
  \centering
  \vspace{-0.1in}
  \begin{subfigure}[b]{0.49\linewidth}
    \includegraphics[width=0.99\linewidth]{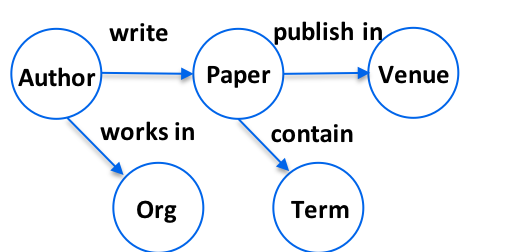}
    \caption{DBLP}
  \end{subfigure}
  \begin{subfigure}[b]{0.49\linewidth}
    \includegraphics[width=0.99\linewidth]{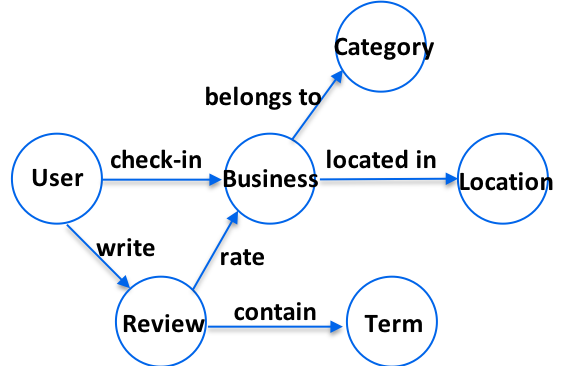}
    \caption{Yelp}
  \end{subfigure}
  \vspace{-0.1in}
  \caption{Examples of heterogeneous network schema}
  \label{fig:egschema}
\end{figure}

Examples of heterogeneous information network include the popular DBLP bibliographic network and Yelp social information network. Figure~\ref{fig:egschema} presents the network schema, i.e., meta-template for an information network, for each of the example instances. In Figure~\ref{fig:egschema}(a), multiple types of objects such as authors, papers, conference venues, author organizations, and paper keywords are connected by multiple types of relationships such as authorship (author $\rightarrow$ paper), affiliation (author $\rightarrow$ organization), etc. In Figure~\ref{fig:egschema}(b), multiple types of objects such as users, businesses, business locations, user reviews, and review terms are connected by multiple types of relationships such as check-in (user $\rightarrow$ business), etc.     

We define our representation learning task on such a heterogeneous network.

\begin{defn}{\textbf{Heterogeneous Network Representation Learning:}} Given a heterogeneous network G, the goal of representation learning is to learn a function f: V $\rightarrow$ $\mathbb{R}^d$ that maps nodes in G to d-dimensional features in vector space and learns $X \in \mathbb{R}^{|V| \times d}$, d $\ll$ $|V|$ such that network structure and semantic heterogeneity is preserved.     
\end{defn}

We leverage motifs to design our heterogeneous network representation learning method.   

\begin{defn}{\textbf{Heterogeneous Network Motif:}} 
A network motif $M$ = ($V_M$, $E_M$, $T_{V_M}$, $T_{E_M}$) is an isomorphic induced directed subgraph consisting of a subset of k nodes from directed heterogeneous network G with $V_M\in V$, $E_M \in E$, $T_{V_M} \in T_V$, $T_{E_M} \in T_E$, 
such that: 

\noindent
(i) $|V_M| = k$, 

\noindent
(ii) $E_M$ consists of all of the edges in $E$ that have both endpoints in $V_M$,

\noindent
(iii) $(u, v) \in E_M$ iff  $(f(u), f(v)) \in E$ for mapping function $g$: $V_M \rightarrow V$, and 

\noindent
(iv) frequency $F_{M_k}$ of appearance of $M$ in $G$ is above a predefined threshold (i.e., statistically significant).  
\end{defn}

A recurring pattern is considered statistically significant if the frequency of its appearance in a graph is significantly higher than the frequency of its appearance in any randomized network. 



\begin{figure}[t]
  \centering
  \begin{subfigure}[b]{0.60\linewidth}
    \includegraphics[width=0.91\linewidth]{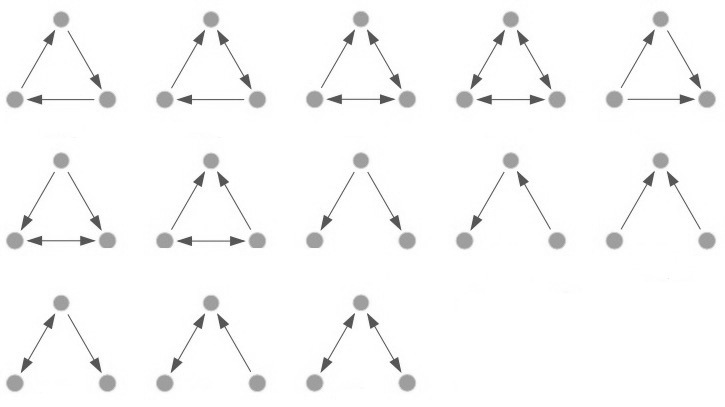}
    \caption{All 3-node network motifs}
  \end{subfigure}
  \begin{subfigure}[b]{0.36\linewidth}
    \includegraphics[width=0.93\linewidth]{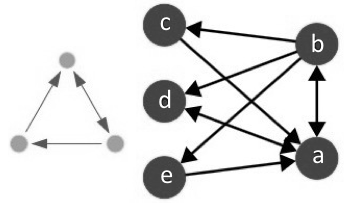}
    \caption{Toy example}
  \end{subfigure}
  \caption{Introduction to motifs}
  \label{fig:motifillustration}
\end{figure}

Motifs are one of the most common higher-order data structures for understanding complex networks and have been popularly recognized as fundamental units of network~\cite{DBLP:journals/science/BensonGL16}. It has been successfully used in many network mining tasks such as clustering~\cite{DBLP:conf/www/TsourakakisPM17}, anomaly detection (densest subgraph sparsifiers)~\cite{DBLP:conf/allerton/Tsourakakis16}, and convolution~\cite{DBLP:journals/corr/abs-1711-05697}. In this work, we focus on directed motifs and directed heterogeneous network since they offer greater scope of representing rich semantics. Figure~\ref{fig:motifillustration}(a) presents all possible 3-node network motifs. Figure~\ref{fig:motifillustration}(b) is a toy example showing how to find motif instance(s) in a graph~\cite{DBLP:journals/science/BensonGL16}.

In the toy example Figure~\ref{fig:motifillustration}(b), Figure~\ref{fig:motifillustration}(b)(right) depicts the graph and Figure~\ref{fig:motifillustration}(b)(left) depicts the motif. We observe that there are two instances of the motif in the graph: (i) (\{a, b, c\}, \{a, b\}) and (ii)(\{a, b, e\}, \{a, b\}). The instance (\{a, b, d\}, \{a, b\}) is not included as an instance because the induced subgraph on the nodes a, b, and d is not isomorphic to the original graph. 

Motifs are distinctly different from some of the other popular graph meta-structures such as metapath and metagraph. We discuss this in details in Section~\ref{sec:expt}.

\vspace{0.10in}
\noindent
\fbox{\begin{minipage}{26em}
\begin{problem}
Given a directed unweighted heterogeneous information network and a set of network motifs, learn low-dimensional latent representations for each node in the network such that the higher-order heterogeneous network neighborhood structure and semantics is preserved.    
\end{problem}
\end{minipage}}

\section{The motif2vec Framework}
\label{sec:framework}

We present our general motif2vec framework that learns high quality node embeddings for heterogeneous networks. Our approach returns representations that help maximize the likelihood of preserving network neighborhoods of multi-type nodes and links. 

\vspace{0.05in}
\noindent
\textbf{Skip-gram Model:}
First, we introduce word2vec~\cite{DBLP:journals/corr/abs-1301-3781}\cite{DBLP:conf/nips/MikolovSCCD13} and discuss its application to network embedding generation tasks. Mikolov {\em et al.} introduced word2vec group of models that learns the distributed representations of words in a corpus. Specifically, the skip-gram model learns high-quality vector representations of words from large amounts of unstructured text data. The algorithm scans the words in a document and aims to embed every word such that the word’s features can predict nearby context words. 
Deepwalk~\cite{DBLP:conf/kdd/PerozziAS14} and node2vec~\cite{DBLP:conf/kdd/GroverL16} generalized the idea to a homogeneous network 
by converting a network into a ordered sequence of nodes. For this, both methods sample sequences of nodes from the original network by random walk strategies.

\vspace{0.05in}
\noindent
\textbf{Random Walk:} A walk in a graph or directed graph $G = \{V, E\}$ is a sequence of nodes  $v_1, v_2, ..., v_{k+1}$ ($v_i \in V$), not necessarily distinct, such that $(v_i,v_{i+1}) \in E$. When the consecutive nodes in the sequence are selected at random, we generate a a random sequence of nodes known as the random walk on the graph. 
A random walk on a graph is a special case of a Markov chain that is time-reversible. The probability of transition from node $v_i$ to $v_j$ is a function of the out-degree of node $v_i$. We explore the neighborhood of a node in a graph or a directed graph using random walk. Specifically, we employ biased random walk procedure that efficiently explores nodes' diverse neighborhoods in both breadth-first and depth-first search fashion~\cite{DBLP:conf/kdd/GroverL16}.   

Such a random walk combined with skip-gram based embedding method learns feature representations $f(u)$ for node $u$ in a homogeneous graph $G' = \{V_{G'}, E_{G'} \}$ that predicts node $u'$s context neighborhood $N(u)$.  

\vspace{-0.05in}
\begin{equation}
max \sum_{u \in V_{G'}} \log \ \Pr (N(u) | f(u))
\end{equation}

\vspace{-0.1in}
Unlike all previous efforts belonging to this family of node embedding algorithms that employ random walk on the original graph~\cite{DBLP:conf/kdd/DongCS17}\cite{DBLP:conf/kdd/GroverL16}\cite{DBLP:conf/kdd/PerozziAS14}\cite{DBLP:conf/www/TangQWZYM15}\cite{DBLP:conf/pakdd/ZhangYZZ18}, we conduct random walk on a transformed graph, known as the motif graph.

\vspace{0.05in}
\noindent
\textbf{Motif Graph:} The network motif literature has defined several graph features and concepts for motifs such as motif cut, motif volume, motif conductance, etc.~\cite{DBLP:journals/science/BensonGL16}. We present one of them, namely motif graph or motif adjacency matrix, which has been used in our algorithmic framework. Given a directed heterogeneous network G = (V, E, $T_V$, $T_E$) and a motif set $M =  \{M_1, M_2, ..., M_T\}$, we compute the motif adjacency matrices $\{W_{M_1}, W_{M_2}, ..., W_{M_T}\}$. The weighted motif adjacency matrix for motif $M_t$ is defined as:

\vspace{-0.2in}
\begin{align}
(W_{M_t})_{ij} &= number\ of\ motif\ instances\ in\ M_t \in M\ where \nonumber \\
           & nodes\ i\ and\ j\ (i \neq j)\ participate\ in\ M_t%
\end{align}

\vspace{-0.1in}
The motif adjacency matrix, also known as the motif co-occurrence matrix, differs from the original graph structurally. The motif graph captures pairwise relationships between nodes in the original graph with respect to a motif. The larger the value in $(W_{M_t})_{ij}$ is, the more significant the relation between nodes $i$ and $j$ is with respect to the motif $M_t$. The motif adjacency matrix can be both weighted or binary. In the later case, $(W_{M_t})_{ij}$ is either 1 or 0 indicating the existence of a relationship between nodes $i$ and $j$ for motif $M_t$. The motif adjacency matrix is symmetric, and thereby undirected. All edges in the original graph may not exist in the motif graph since a motif may not appear for a given edge. The edges in a motif graph are likely to have different weights than the original graph since a motif may appear at a different frequency than another random motif for a given edge. Thus, the number of edges in a weighted motif graph is usually greater than the number of edges in the original graph.

\vspace{0.10in}
We transform the original graph to a motif graph in order to simultaneously encode the heterogeneity in structure and semantics, and conduct random walks on the motif graph itself. 
Additionally, we conduct random walks on the original graph  
to ensure greater coverage of higher-order connected nodes, which may otherwise be missed due to their non-participation in popular motifs. This strategy enables our random walk to be not dependent on the type of the node or link, as in prior art for heterogeneous networks~\cite{DBLP:conf/kdd/DongCS17}\cite{DBLP:conf/pakdd/ZhangYZZ18}. Note that, meta-structure (metapath, metagraph, etc.) driven random walks limit the scope of a walk to explore higher-order diverse neighborhoods. The generated walk sequences are aggregated and shuffled before being fed to skip-gram. Our graph transformation followed by graph meta-structure independent biased random walk enable the sequences to carry both higher-order heterogeneous network structural patterns as well as heterogeneous semantic relationships. We demonstrate the superiority our novel idea empirically in Section~\ref{sec:expt}.  

\begin{figure*}[t]
  \centering
    \includegraphics[width=0.98\linewidth]{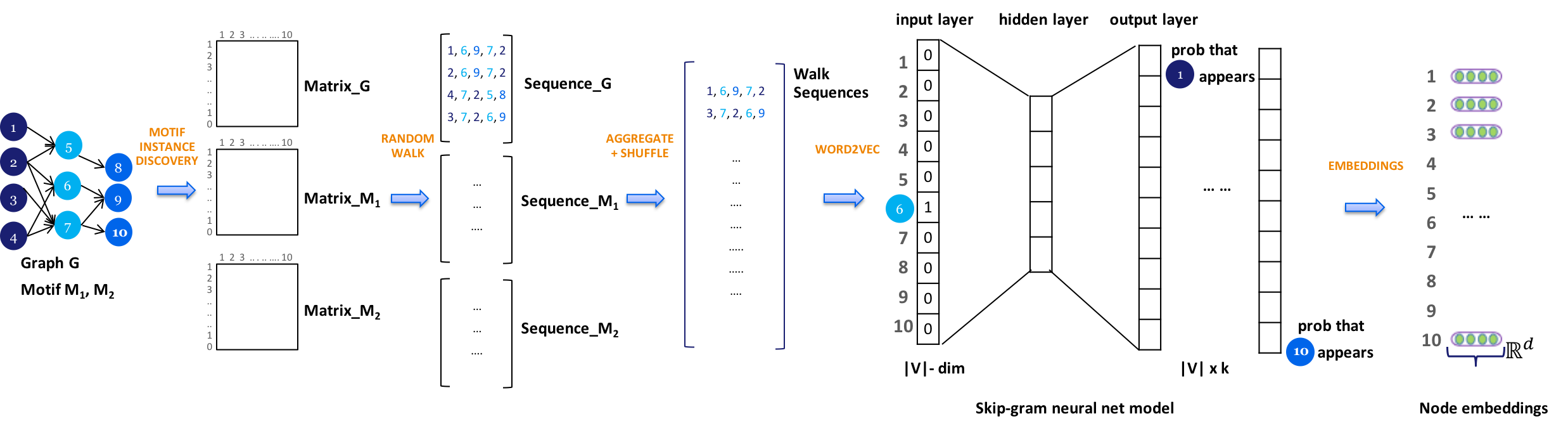}
\vspace{-0.10in}
  \caption{The motif2vec framework}
  \label{fig:framework}
  \vspace{-0.10in}
\end{figure*}

Figure~\ref{fig:framework} illustrates our motif2vec framework. Given a heterogeneous information network $G$ and a set of motifs $M = \{M_1, M_2\}$, the goal of the framework is to output $d$-dimensional embedding vectors for each node in $G$. The steps in the framework presented in Figure~\ref{fig:framework} includes: motif instance discovery, random walk sequence generation, aggregation and shuffling of the generated walk sequences, skip-gram neural net training, and finally embedding generation. We present the three phases of our framework next. 

\begin{algorithm}[t]  
\caption{The motif2vec algorithm}  
\label{alg:motif2vec}  
\begin{flushleft}
\textbf{INPUT:} Heterogeneous information network $G = (V, E,$ $T_V, T_E)$, motif set $M = \{M_1, ... M_T\}$, embedding dimension $d$, walks per node $r$, walk length $l$, neighborhood size $k$, return parameter $p$, in-out parameter $q$\\
\textbf{OUTPUT:}  Latent node representations $X \in R^{|V| \times d}$
\end{flushleft}
\begin{algorithmic}[1]

\State  Initialize $X$
\State  Initialize $\Psi$
\State $\Psi \leftarrow$ Discover-Motif-Instances($G$, $M$)		    
\State Initialize $W_M$
\For {$t=1$ to $T$}
   \State $W_{M_t} \leftarrow$ Create-Weighted-Motif-Graph ($G$, $M_t$, $\Psi$)  
\EndFor 
\State Initialize {\em walks}
\For {$G$ and $W_{M_1}, W_{M_2}, ..., W_{M_T}$}
\ForAll {nodes $u \in V_{G'}$} // \small current iter $G' = \{V_{G'}, E_{G'}\}$ \normalsize
   \State $walk$ $\leftarrow$ Generate-Random-Walk $(G', u', l)$ 
   \State Append $walk$ to $walks$
\EndFor
\EndFor
\State Initialize {\em sequences}
\State $sequences$ $\leftarrow$ Shuffle ($walks$)
\State $X$ $\leftarrow$ Skip-Gram-Model($sequences$, $k$, $d$)
\State Return $X$ 
\end{algorithmic}
\end{algorithm}

\vspace{0.05in}
\noindent
\textbf{Network Transformation}: First, we find instances of the motif(s) under consideration in the original network. This is referred to as the motif discovery task in the literature and is a computationally expensive operation. Many motif discovery algorithms have been proposed over the years, each with the intent of improving the computational aspects of the state-of-the-art~\cite{DBLP:journals/ijdmb/Kavurucu15}. We use the method presented in~\cite{DBLP:conf/scipy/HagbergSS08} for motif discovery. Once the motif instances are received, we compute the weighted motif adjacency matrix. Thus, we transform the original network to motif graph(s) that encodes the heterogeneity in network structure and semantics.

\vspace{0.05in}
\noindent
\textbf{Sequence Generation}: Next, we generate random walk sequences for motif graph(s) and the original graph. We generate random walks on both transformed graph(s) and original graph. We use the method in~\cite{DBLP:conf/kdd/GroverL16} for generating sequences. We aggregate and shuffle the sequences generated from the original and the motif graph(s) before feeding them to the neural net. Thus, our generated walk sequences encompasses both local and global network heterogeneous connectivity. 

\vspace{0.05in}
\noindent
\textbf{Embedding Generation}: Finally, we input the walk sequences from the previous step and output node embeddings. We use the skip-gram neural net model architecture in ~\cite{rehurek_lrec} for learning the latent feature representations. We minimize our optimization function using SGD with negative sampling that is known to learn accurate representations efficiently~\cite{DBLP:conf/nips/MikolovSCCD13}. Following Equation 1, we optimize embedding $z_u$ ($z_u \in \mathbb{R}^d$) for node $u$  in graph $G$ for random walk co-occurrences according to: 
\begin{align}
\mathcal{L} &=  \sum_{u \in V} \sum_{v \in N(u)} -\log (Pr(v|z_u)) \nonumber \\
&= \sum_{u \in V} \sum_{v \in N(u)} -\log (\frac{\exp(z^T_u z_v)}{\sum_{n \in V}\exp(z^T_u z_n)})  
\end{align}  
where $N(u)$ is the neighborhood of node $u$ and node $v \in N(u)$ is seen on a random walk staring from node $u$.

\noindent 
The pseudo-code for motif2vec is presented in Algorithm~\ref{alg:motif2vec}.


\section{Experiments}
\label{sec:expt}

We evaluate the heterogeneous network node embeddings obtained through motif2vec on two standard supervised machine learning tasks: multi-label node classification and link prediction.   

\vspace{-0.05in}
\subsection{Experimental Setup}

We compare motif2vec with several recent network representation learning algorithms on multiple datasets.

\vspace{0.01in}
\subsubsection{Datasets}
We use three popular publicly available heterogeneous networks data from the literature: 

\vspace{0.02in}
\noindent
\textbf{DBLP-P Dataset}: It is a bibliographic network composed of three types of nodes: author (A), paper (P), and venue (V) connected by three types of links: A $\rightarrow$ P, P $\rightarrow$ V, and P $\rightarrow$ P. We use a subset of the DBLP
dataset made available by ~\cite{DBLP:journals/corr/abs-1711-05697}\cite{DBLP:journals/pvldb/SunHYYW11} for paper classification task. The papers are labeled to belong to 10 classes such as {\em information retrieval}, {\em databases}, {\em networking}, {\em artificial intelligence}, {\em operating systems}, etc. that are extracted from Cora~\cite{DBLP:journals/ir/McCallumNRS00}. There are 17,411 authors (A), 18,059 papers (P), and 300 conferences, i.e., venues (V). 
 
\vspace{0.01in}
\noindent
\textbf{AMiner-CS Dataset}: It is another bibliographic network graph composed of three types of nodes: author (A), paper (P), and venue (V) connected by three types of links: A $\rightarrow$ P, P $\rightarrow$ V, and P $\rightarrow$ P. We use a version of the AMiner
Computer Science (CS) dataset made available by~\cite{DBLP:conf/kdd/DongCS17} for author classification. It comprises of 1,693,531 authors (A), 3,194,405 papers (P), and 3,883 venues (V). Author research categories are labeled to belong to 8 classes such as {\em theoretical computer science}, {\em computer graphics}, {\em human computer interaction}, {\em computer vision and pattern recognition}, 
etc. based on the categories in Google Scholar~\cite{DBLP:conf/kdd/DongCS17}. There are 246,678 labeled authors in this dataset. We use it for author node classification task.  

\begin{figure}[t]
  \centering
    \includegraphics[width=0.96\linewidth]{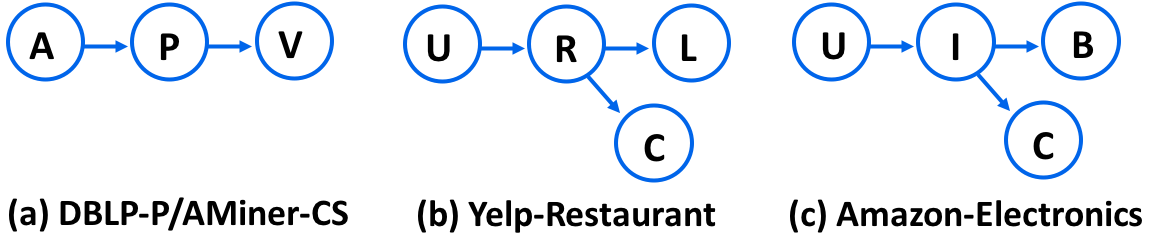}
    \vspace{-0.00in}
  \caption{Heterogeneous information network schemas}
  \label{fig:dataschema}
   \vspace{-0.15in}
\end{figure} 


\vspace{0.03in}
\noindent
\textbf{ 
Yelp-Restaurant Dataset}: We consider the data obtained from the 12$^{th}$ round of Yelp Dataset Challenge.
We build a heterogeneous network composed  of four types of nodes: users (U), businesses, i.e., restaurants (R), location (L), and category (C) connected by three types of links: U $\rightarrow$ R, R $\rightarrow$ C, and C $\rightarrow$ L. Yelp dataset includes users who have very few reviews. In fact, about 49\% of the users have only one review~\cite{DBLP:conf/sigir/ZhangL0ZLM14} making the dataset very sparse and hence difficult for evaluation purposes. Following the common practice by other works (e.g.,~\cite{DBLP:conf/sigir/ZhangL0ZLM14}), we filter out users with less than twenty business reviews over fourteen years (2004 - 2018). There are 36,432 users (U), 18,256 restaurants (R), 5,514 locations, and 419 categories (C). We use this data for U $\rightarrow$ R link prediction task. 

\vspace{0.03in}
\noindent
\textbf{Amazon-Electronics Dataset}: We consider the Amazon-200k data-set~\cite{DBLP:conf/www/HeM16}\cite{DBLP:conf/kdd/ZhaoYLSL17} which contains ratings provided by users on electronics items in Amazon.
Similar to Yelp-Restaurant dataset, we build a heterogeneous network composed of four types of nodes: users (U), items (I), brand (B) and category (C) connected by 3 types of links: U $\rightarrow$ I, I $\rightarrow$ B, and I $\rightarrow$ C. There are 59,297 users (U), 21,000 items (I), 2,059 brands (B), and 683 categories (C). We use this dataset for U $\rightarrow$ I link prediction.

\begin{table}[t]
\begin{center}
\begin{tabular}{| c | c | c|}
\hline
 Dataset & \#Nodes & \#Links \\ \hline
 DBLP-P & 35,770 & 131,636 \\  
 AMiner-CS & 4,891,819 & 12,506,615\\
 Yelp-Restaurant & 60,621 & 189,423\\
 Amazon-Electronics & 83,039 & 284,650 \\ \hline
\end{tabular}
\caption{Heterogeneous information network statistics}
\vspace{-0.10in}
\label{tbl:datastats}
\end{center}
\vspace{-0.20in}
\end{table}

\begin{table*}[t]
\begin{center}
\begin{tabular}{|*{5}{>{\centering\arraybackslash}p{0.17\linewidth}|}}
\hline
\multirow{2}{*}{Method} & \multicolumn{2}{c|}{Multi-Class Node Classification} & \multicolumn{2}{c|}{Link Prediction}\\
\cline{2-5}
 & DBLP-P & AMiner-CS & Yelp-Restaurant & Amazon-Electronics\\
\hline
 motif2vec & 78.80 & 91.68 & 58.38 & 58.90 \\
\hline
metapath2vec & 60.08 & 73.90 & 43.30 & 50.89 \\
\hline
metapath2vec++ & 49.40 & 72.31 & 29.21 & 57.02 \\
\hline
metagraph2vec & 64.48 & 82.09 & 29.24 & 55.53 \\
\hline
metagraph2vec++ & 53.24 & 35.58 & 39.60 & 60.02 \\
\hline
\end{tabular}
\caption{Quantitative results (accuracy in \%) for different machine learning tasks, different datasets, different embedding methods under the same experimental settings}
\label{tbl:results}
\vspace{-0.10in}
\end{center}
\end{table*}

We build heterogeneous information networks out of each of the datasets. The network schema and statistics are presented in Figure~\ref{fig:dataschema} and Table~\ref{tbl:datastats} respectively. 

\subsubsection{Baseline Methods}
We compare motif2vec with {\em recent} network representation learning methods focused on heterogeneous networks. Specifically, we focus on the family of node embedding methods to which motif2vec belongs.    

\vspace{0.03in}
\noindent
\textbf{metapath2vec, metapath2vec++~\cite{DBLP:conf/kdd/DongCS17}}: Dong \textit{et al.} study the problem of representation learning in heterogeneous networks. They propose two models: metapath2vec and metapath2vec++ that first leverages meta-path based random walks to construct the heterogeneous neighborhood of a node and then leverages a heterogeneous skip-gram model to generate the embeddings.      
  
\vspace{0.03in}
\noindent
\textbf{metagraph2vec, metagraph2vec++~\cite{DBLP:conf/pakdd/ZhangYZZ18}}: Zhang \textit{et al.} proposed a network embedding learning method for heterogeneous networks that leverages metagraph to capture richer structural contexts and semantics between distant nodes. The method uses metagraph to guide the generation of random walks and then skip-gram model to learn latent embeddings of multi-typed heterogeneous network nodes. metagraph2vec uses homogeneous skip-gram model while metagraph2vec++ uses heterogeneous skip-gram model.     

The authors in~\cite{DBLP:conf/kdd/DongCS17} demonstrate how the proposed method beats some of the popular state-of-art at that time, namely DeepWalk~\cite{DBLP:conf/kdd/PerozziAS14}, node2vec~\cite{DBLP:conf/kdd/GroverL16}, LINE~\cite{DBLP:conf/www/TangQWZYM15}, Spectral Clustering~\cite{DBLP:conf/dmkd/TangL11} and Graph Factorization~\cite{DBLP:conf/www/AhmedSNJS13}. Thus, we exclude them from our experiments. Note that, each of this work considers homogeneous networks.

\subsubsection{Machine Learning Tasks} We consider two standard supervised machine learning tasks to evaluate our embedding. 

\vspace{0.03in}
\noindent
\textbf{Node Classification:} Node classification is a downstream machine learning task that 
classifies nodes in a heterogeneous network into a pre-defined set of classes. We follow a standard classification setup and use the generated node embeddings as features for the classifier. We conduct paper node multi-class classification for DBLP-P data and author node classification for AMiner-CS data.    
The classifier, parameter values, and train/test data is fixed for the various embedding approaches to avoid any confounding factor. We use traditional SVM classifier for both DBLP-P and 
AMiner-CS datasets, without any parameter tuning. 

\vspace{0.03in}
\noindent
\textbf{Link Prediction:}
Link prediction is a widely popular machine learning task in heterogeneous networks that predicts links that are likely to be added to the network in the near future. 
We leverage node embedding features to predict links. We partition the links in a network to train and test instances in order to hide a fraction of the existing links during embedding learning. The probability of a link appearing between two nodes in a network is calculated by computing cosine similarity between the respective feature vector embeddings. In our experiments, if the embedding-based similarity score between a pair of nodes is higher than a threshold, we infer that an edge could exist between the two nodes. In order to penalize embeddings that generate a high similarity value for any random pair of nodes, we generate an equal number of fake links in the test set. These fake links correspond to links that do not exist in the original network. The intuition is that these embeddings are expected to return a similarity score less than the threshold. 
We evaluate our embeddings for link prediction on Yelp-Restaurant and Amazon-Electronics datasets.   


\subsubsection{Evaluation Metric}
In traditional classification task, accuracy is a popular evaluation metric and we consider that. We perform the standard 70:30 split for train and test data, and measure the percentage of correct predictions for the test instances during multi-class classification. For link prediction, we split data into 70:30 such that the links present in the test set are removed from the original network on which embeddings are learnt. We measure the percentage of correct predictions, i.e., presence of links, for the test instances. We refer to our link prediction evaluation metric as accuracy too.

\subsubsection{Settings}
\label{settings}

For all embedding methods, we use the exact same parameters listed below. The parameter settings used are in line with typical values used in prior art~\cite{DBLP:conf/kdd/GroverL16}.

$-$  The embedding vector dimension $d$: 128

$-$  The walk length $l$: 80

$-$  The number of walks per node $r$: 10

$-$  The context size for optimization $c$: 10 

$-$  Random walk return parameter $p$: 1 

$-$  Random walk in-out parameter $q$: 1  

\noindent
The optimization is run for a single epoch. Each of our reported numbers is an average of five runs. All codes are implemented in Python All experiments are conducted on a Linux machine with 2.60GHz Intel processor, 28 CPU cores, and 800GB RAM. 


\begin{figure*}[t]
  \centering
    \includegraphics[width=0.96\linewidth]{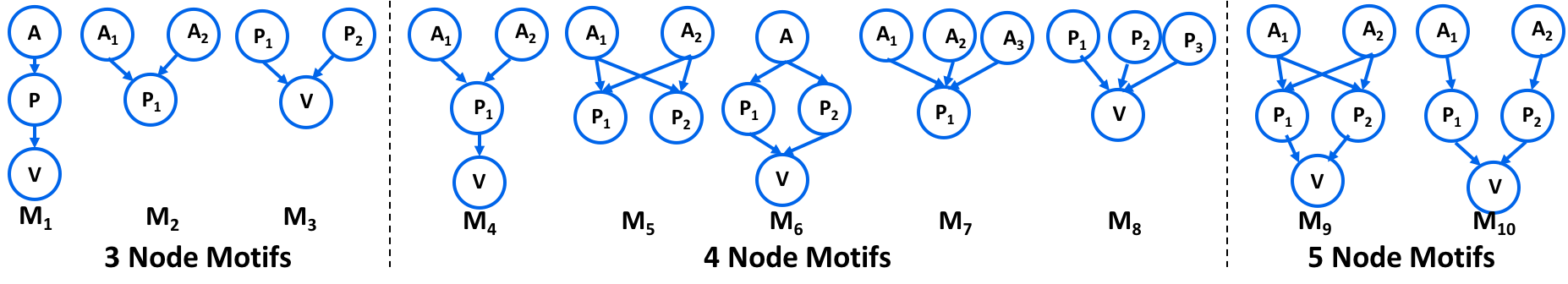}
    \vspace{-0.10in}    
  \caption{Example 3-node, 4-node, and 5-node motifs for DBLP-P/AMiner-CS datasets for the heterogeneous network schema in Figure~\ref{fig:dataschema}(a)}
  \label{fig:dblp-mo}
\end{figure*}

\begin{figure}[t]
  \centering
  \vspace{-0.20in}
    \includegraphics[width=0.84\linewidth]{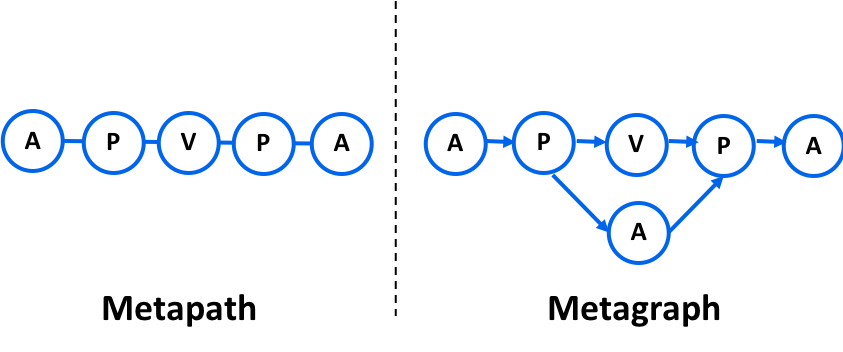}
      \vspace{-0.10in}
  \caption{metapath and metagraph for DBLP-P/AMiner-CS datasets for the heterogeneous network schema in Figure~\ref{fig:dataschema}(a)}
  \label{fig:dblp-mpg}
    \vspace{-0.10in}
\end{figure}

\subsection{Results: Accuracy}
Table~\ref{tbl:results} presents our experimental results for the different machine learning tasks: multi-class node classification and link prediction, different datasets: DBLP-P, AMiner-CS, Yelp-Restaurants, and Amazon-Electronics, different embedding methods: motif2vec, metapath2vec, metapath2vec++, metagraph2vec, and metagraph2vec++ under the same configuration parameter settings, as detailed in Section 4.1. We observe that our algorithm motif2vec consistently and significantly outperforms the baseline methods for both tasks and across all four datasets.

For paper node classification task on DBLP-P data, motif2vec beats the best baseline by 22\%. For author research category classification task on AMiner-CS data, motif2vec beats the best baseline by 24\%. For $U \rightarrow R$ link prediction task on Yelp-Restaurant data, motif2vec achieves 34\% improvement, while for $U \rightarrow B$ link prediction on Amazon-Electronics data, motif2vec achieves 3\% improvement over the second best baseline method. We observe that metagraph2vec++ is the best algorithm for Amazon-Electronics dataset. However, metagraph2vec and metagraph2vec++ are fairly inconsistent, as is evident from their accuracy numbers for the remaining datasets in the table. The authors in \cite{DBLP:conf/kdd/DongCS17} and \cite{DBLP:conf/pakdd/ZhangYZZ18} introduce the ``++'' version of metapath2vec and metagraph2vec since the heterogeneous skip-gram model is expected to accommodate the heterogeneity in network better. However, the results presented by the authors in both works fail to showcase the steady benefits of heterogeneous skip-gram. We do not propose an extended version of motif2vec in this paper.        

\begin{table}[t]
\begin{center}
\vspace{-0.20in}
\begin{tabular}{| c | c |}
\hline
 Motif & Accuracy (in \%) \\ 
 \hline
 A 3-node motif ($M_2$) & 78.50 \\
 \hline
 A 4-node motif ($M_4$) & 78.80 \\
 \hline
 A 5-node motif ($M_{10}$) & 78.43 \\
 \hline
 All 3-node motifs ($M_1, M_2, M_3$) & 78.00 \\
 \hline
 All 4-node motifs ($M_4, M_5, M_6, M_7, M_8$) & 77.75 \\
 \hline
\end{tabular}
\caption{Multi-Class Paper Node Classification for DBLP-P}
\vspace{-0.10in}
\label{tbl:diffmotifs}
\end{center}
\vspace{-0.20in}
\end{table}

In summary, our method learns consistently and significantly better (achieving relative improvements as high as 24\% and 34\% over benchmarks for classification and prediction respectively) heterogeneous network node embeddings than existing state-of-the-art methods. This is primarily because transforming the original complex graph to a motif graph helps accommodate heterogeneous network structure and semantic heterogeneity effectively. 




\begin{figure*}[h]
  \centering
  \begin{subfigure}[b]{0.49\linewidth}
    \includegraphics[width=0.99\linewidth]{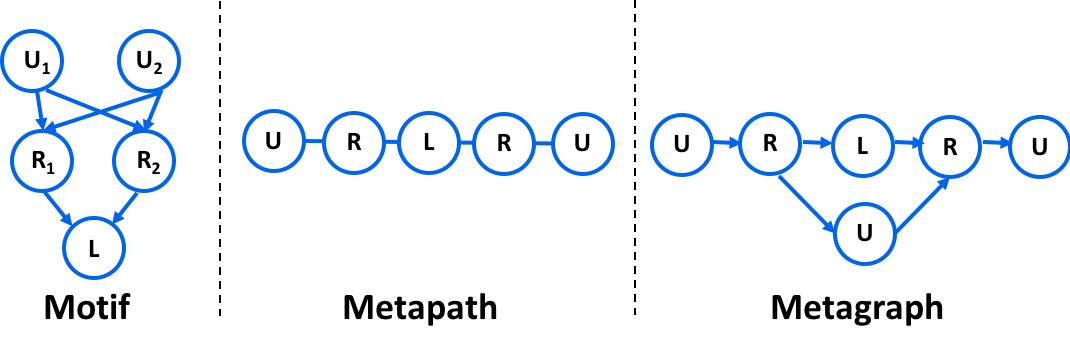}
    \vspace{-0.10in}
    \caption{Yelp-Restaurant}
  \end{subfigure}
  \hfill
  \begin{subfigure}[b]{0.49\linewidth}
    \includegraphics[width=0.99\linewidth]{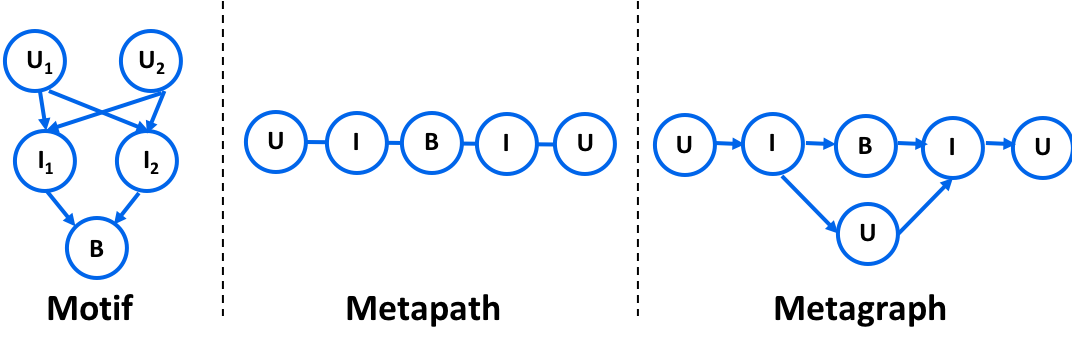}
    \vspace{-0.10in}
    \caption{Amazon-Electronics}
  \end{subfigure}
    \vspace{-0.01in}
  \caption{motif, metapath, and metagraph for Yelp-Restaurant and Amazon-Electronics datasets for the heterogeneous network schemas in Figure~\ref{fig:dataschema}(b) and Figure~\ref{fig:dataschema}(c) respectively}
  \label{fig:yelp-amazon-mmm}
  \vspace{-0.18in}
\end{figure*}

\vspace{-0.10in}
\subsection{motif vs. metapath vs. metagraph}
Figure~\ref{fig:dblp-mo} illustrates example 3-node, 4-node, and 5-node motifs for the machine learning task on DBLP-P and AMiner-CS datasets. Figure~\ref{fig:dblp-mpg} presents the metapath and metagraph for node classification task on DBLP-P and AMiner-CS datasets. 

Motifs are crucial to understanding the structure, semantics, and functions of meaningful patterns in complex networks. Thus, interesting motifs in a heterogeneous bibliographic network may include (motifs in Figure~\ref{fig:dblp-mo}): (i) authors collaborating on the same paper ($M_2$ and $M_7$), (ii) papers published at the same venue ($M_3$ and $M_8$), (iii) authors collaborating on a paper that gets published at a venue ($M_4$), (iv) authors collaborating on papers ($M_5$), (v) author publishing papers at the same venue ($M_6$), (vi) authors collaborating on papers that get published at a venue ($M_9$), and (vii) authors publishing papers at the same venue ($M_{10}$).

Table~\ref{tbl:diffmotifs} presents the effectiveness of each motif in generating higher quality embeddings useful for classification. In our framework, we can combine multiple motifs for learning more effective node representations. Some combinations of motifs are more useful than the others. In Table~\ref{tbl:diffmotifs}, all 3-node motifs and all 4-node motifs do not return the highest classification accuracy since at least one non-useful motif in each set pulls the classification accuracy down. Determining the  best combination of motifs for higher quality embedding learning is combinatorially expensive and not the focus of this paper. In our experiments, we consider only one motif, i.e., $M_4$ in order to ensure a fair comparison with the baseline methods which consider one metapath and one metagraph with the same semnatics (as $M_4$) respectively.  

Authors in~\cite{DBLP:conf/kdd/DongCS17} surveyed metapath related efforts and found that the most popularly used meta-path schemes in bibliographics networks are $A-P-A$ and $A-P-V-P-A$. $A-P-A$ denotes co-authorship relationship while $A-P-V-P-A$ represents authors publishing papers at the same venue. We consider $A-P-V-P-A$ as the metapath for experiments involving DBLP-P and AMiner-CS dataset since it can be generalized to diverse tasks in a heterogeneous bibliographic network~\cite{DBLP:conf/kdd/DongCS17}.
Authors in~\cite{DBLP:conf/pakdd/ZhangYZZ18} extend metapaths to metagraphs in order to capture rich contexts and semantic relations between nodes better. The augmentation of $A \rightarrow P \rightarrow A \rightarrow P \rightarrow A$ path to the directed metapath $A \rightarrow P \rightarrow V \rightarrow P \rightarrow A$ helps the meta-structure encode semantic relations between distant nodes. Thus, we choose the same metagraph as the one shown in~\cite{DBLP:conf/pakdd/ZhangYZZ18} for heterogeneous bibliographic network mining.

Figure~\ref{fig:yelp-amazon-mmm}(a) and Figure~\ref{fig:yelp-amazon-mmm}(b) present the motif, metapath, and metagraph for link prediction on Yelp-Restaurant and Amazon-Electronics datasets respectively. Our choice of metapath and metagraph for Yelp-Restaurant and Amazon-Electronics datasets is inspired by~\cite{DBLP:conf/kdd/ZhaoYLSL17}. 
Similar to our set-up for DBLP-P and Aminer-CS datasets, we choose one motif from the set of possible motifs for motif2vec in order to ensure a fair comparison with metapath2vec and metagraph2vec. Note that, each of our motif, metapath, and metagraph in Figure~\ref{fig:yelp-amazon-mmm}(a) and Figure~\ref{fig:yelp-amazon-mmm}(b) consist of three node types though the original schema has four node types. This is because metapath and metagraph cannot handle four types of nodes, as discussed next. Experimental results replacing node type location (L) with node type category (C) and node type brand (B) with node type category (C) for Yelp-Restaurant and Amazon-Electronics dataset respectively are similar to the results in Table~\ref{tbl:results}. 

\begin{figure}[h]
  \centering
  \vspace{-0.10in}
    \includegraphics[width=0.42\linewidth]{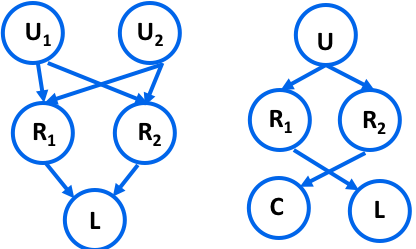}
    \vspace{-0.10in}
  \caption{Example bifan motifs in Yelp-Restaurant dataset for the heterogeneous network schema in Figure~\ref{fig:dataschema}(b)}
  \label{fig:bifan}
\end{figure}

\noindent
\textbf{Advantages of motif over metapath and metagraph:} Motifs are capable of capturing greater context and leveraging richer semantics than both metapaths and metagraphs. This is because both metapaths and metagraphs are commonly used in a symmetric way thereby facilitating a recursive guidance for random walkers~\cite{DBLP:conf/kdd/DongCS17}\cite{DBLP:journals/tkde/ShiLZSY17}\cite{DBLP:series/synthesis/2012Sun}\cite{DBLP:conf/pakdd/ZhangYZZ18}. Thus, they cannot build meaningful meta-structures for heterogeneous network schemas like Yelp-Restaurant and Amazon-Electronics having four node types. 
Figure~\ref{fig:dblp-mo} reveals how a lot more interesting heterogeneous patterns can be captured by motif than by metapath and metagraph. 
Figure~\ref{fig:bifan} showcases two example {\em bifan} motifs, known to occur frequently in complex networks, for Yelp-Restaurant dataset that a metapath or a metagraph cannot capture. In addition, metapath2vec and metagraph2vec are designed to operate only on a single metapath and a single metagraph respectively, unlike motif2vec.


\vspace{-0.15in}
\subsection{Efficiency}
It is important to investigate the efficiency of motif2vec in today's age of big graph data. The computationally expensive steps in the algorithm are: motif instance discovery, sequence generation, and embedding learning. The motif instance discovery literature is about two decades old and boasts of many efficient algorithms~\cite{DBLP:journals/ijdmb/Kavurucu15}.
In this work, we consider the widely adopted formalization of the motif discovery task, namely subgraph isomorphism, and 
use the fast method presented in ~\cite{DBLP:conf/scipy/HagbergSS08} (NetworkX library) for motif discovery. For sequence generation and embedding learning using skip-gram, we employ the parallelization tricks suggested by prior-art~\cite{DBLP:conf/kdd/GroverL16}\cite{DBLP:conf/nips/MikolovSCCD13}. 

\begin{figure}[h]
  \vspace{-0.10in}
  \centering
    \includegraphics[width=0.90\linewidth]{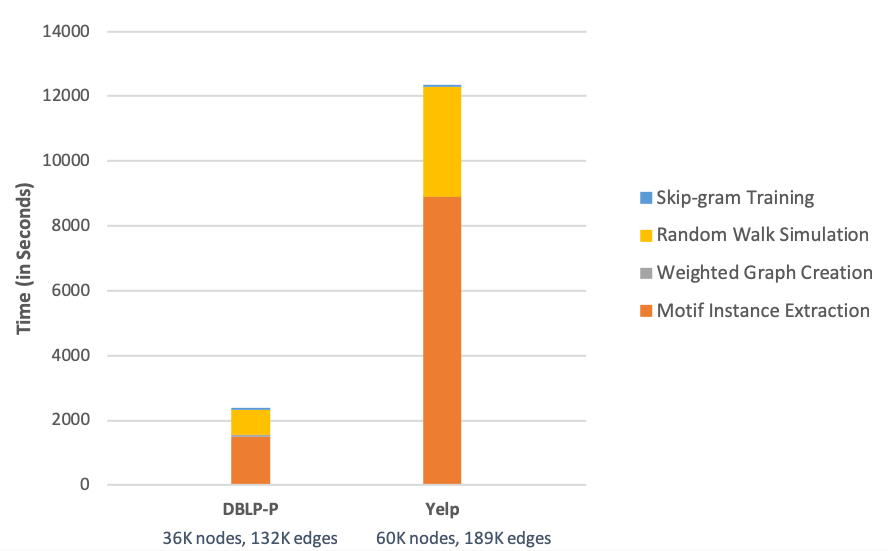}
      \vspace{-0.05in}
  \caption{motif2vec execution time analysis}
  \label{fig:time}
    \vspace{-0.18in}
\end{figure}

Figure~\ref{fig:time} illustrates the time taken by each of the individual steps: motif instance extraction, weighted graph creation, random walk simulation, and skip-gram neural net training for two datasets, one from each machine learning task, under consideration. As expected, the time to extract the motif instances dominate motif2vec algorithm's end-to-end execution time, followed by the time taken to generate random walk sequences. Thus, our algorithm is limited by the scalability of motif instance extraction in the worst case.  

\noindent
\textbf{motif2vec efficiency for 4.9M nodes, 12.5M edge graph:}
We conduct experiments on AMiner-CS dataset heterogeneous network consisting of 4.9M nodes and 12.5 million edges to highlight the effectiveness of our method for big graphs, in spite of the computational expenses associated with motif instance extraction and random walk simulation. Given a heterogeneous network with well-defined semantics and relations (see Figure~\ref{fig:dataschema}) and a motif of interest (see motif $M_4$ in Figure~\ref{fig:dblp-mo}), we implement our own heuristic motif instance extraction method that is guided by the pattern in the motif to identify the matching sub-graphs in the overall network. While NetworkX's module for motif instance extraction returns all possible matching sub-graphs which are far more in number than what is relevant to our task, our heuristic method prunes the candidate space and speeds up the extraction process. We skip the details of our heuristic due to lack of space. 

For Aminer-CS dataset, we run experiments with 24 threads with each of them utilizing a CPU core, under the same parameter settings (Section~\ref{sec:expt}). The random walk sequence generation step uses OpenMP to automatically decide the number of cores, as seen in the high-performance graph analytics SNAP repository\cite{DBLP:conf/kdd/GroverL16}. For Aminer-CS dataset, motif2vec took 8 hours 55 minutes for end-to-end execution, of which:

\noindent $-$ 26 minutes is taken by our heuristic motif instance extraction method,

\noindent $-$ 43 minutes is taken by random walk simulation method,

\noindent $-$ 7 hours 42 minutes is taken by word2vec/skip-gram neural net model training. 

Our algorithm involves a number of parameters, as evident in Algorithm~\ref{alg:motif2vec}. 
We do not conduct parameter sensitivity analysis experiments. Instead, we learn from the authors' efforts in~\cite{DBLP:conf/kdd/DongCS17}\cite{DBLP:conf/kdd/GroverL16} and make our choices.

\vspace{-0.075in}
\section{Related Work}

\noindent
\textbf{Network representation learning:}
Network representation learning is a well-studied research problem owing to the ubiquitous nature of networks in the real-world and applications such as node classification, link prediction, visualization, clustering, etc. 
Various approaches have been studied in the literature to address this problem~\cite{DBLP:journals/kbs/GoyalF18}. 
The early approaches aimed to learn node representations by factorizing the graph adjacency matrix as performed in recommender systems~\cite{DBLP:conf/www/AhmedSNJS13}\cite{DBLP:conf/kdd/OuCPZ016}, and are computationally expensive. 
The random-walk based methods have emerged to be the most popular 
and includes DeepWalk~\cite{DBLP:conf/kdd/PerozziAS14}, node2vec~\cite{DBLP:conf/kdd/GroverL16}, LINE~\cite{DBLP:conf/www/TangQWZYM15}, HARP~\cite{DBLP:conf/aaai/ChenPHS18}, etc. Most of the efforts are designed for homogeneous networks and are inadequate to handle heterogeneous networks. Heterogeneous network representation learning methods include metapath2vec~\cite{DBLP:conf/kdd/DongCS17}, metagraph2vec~\cite{DBLP:conf/pakdd/ZhangYZZ18}, PTE~\cite{DBLP:conf/kdd/TangQM15}, HNE~\cite{DBLP:conf/kdd/ChangHTQAH15}, LANE~\cite{DBLP:conf/wsdm/HuangLH17}, and  ASPEM~\cite{DBLP:conf/sdm/ShiGZK018}. Except metapath2vec and metagraph2vec, none of these methods is aligned to our problem of generic unsupervised task-independent network embedding learning preserving the heterogeneity in structure and semantics, as discussed in Section~\ref{sec:intro}. Of late, deep learning based approaches have become popular for node representation learning. 
Further research focused on interpreting the embedding learned by these models can be useful.  


\noindent
\textbf{Heterogeneous information network:}
Heterogeneous information networks are graphs that have various types of nodes and links fusing more information and containing richer semantics. Heterogeneous information networks are used to model most real-world networks today. 
In the literature, researchers have published various tasks related to heterogeneous networks such as similarity search~\cite{DBLP:journals/pvldb/SunHYYW11}, clustering~\cite{DBLP:conf/kdd/SunNHYYY12}, prediction~\cite{DBLP:conf/wsdm/SunHAC12}, classification~\cite{DBLP:conf/cikm/KongCYDW12}, etc. Each method is designed for a specific heterogeneous network mining application. In our work, we learn node representations that are effective for both classification and prediction. We also demonstrate how motif2vec outperforms existing heterogeneous network representation learning methods~\cite{DBLP:conf/kdd/DongCS17}\cite{DBLP:conf/pakdd/ZhangYZZ18} that cannot capture higher-order heterogeneous connectivity patterns or preserve the structure and semantics of multiple types of nodes/links.    

\noindent
\textbf{Network motifs:}
Network motifs are simple basic building blocks of complex networks. 
Motifs have originated from domains such as biochemistry and ecology where they are used for studying networks such as gene regulation, neuron synaptic connection, etc. It has been successfully used in many computer science network mining tasks such as clustering~\cite{DBLP:conf/www/TsourakakisPM17}, anomaly detection~\cite{DBLP:conf/allerton/Tsourakakis16}, and convolution~\cite{DBLP:journals/corr/abs-1711-05697}.
Rossi {\em et al.} addressed the problem of higher-order network representation learning using motifs for homogeneous networks~\cite{DBLP:conf/www/RossiAK18}. But the method cannot be extended to handle heterogeneous networks. HONE~\cite{DBLP:conf/www/RossiAK18} also does not combine the advantages of random-walk based method and motif-aware method, as we do.  


\vspace{-0.05in}
\section{Conclusion}
In this paper, we study the problem of node representation learning for heterogeneous information networks. We propose a novel efficient algorithm, {\em motif2vec} that leverage higher-order, recurring, and statistically significant network connectivity patterns in the form of network motifs to learn latent representations preserving heterogeneity in network structure and semantics. Unlike existing graph embedding methods for heterogeneous networks that employ some form of graph meta-structure to guide heterogeneous semantics aware random walks through the network, we employ motifs to transform the graph to a motif graph, which in turn, encode the heterogeneity. 
Our method preserves both local and global structural relationships in addition to rich semantic correlations in a network. We empirically demonstrate how the proposed algorithm consistently and significant outperforms state-of-the-art baselines on diverse real-world datasets. 
An important input to our algorithm is the choice of motif(s) from the set of all possible motifs. In the future, we intend to explore the possibility of automatically learning the motif weights for a network or for a task. It will also be interesting to study how our algorithm extends to handle the dynamics of evolving heterogeneous networks.         
\balance

\vspace{-0.08in}
\bibliography{references}{}
\bibliographystyle{plain}

\end{document}